\documentclass[twocolumn,showpacs,prl]{revtex4}
\usepackage{graphicx}%
\usepackage{dcolumn}
\usepackage{amsmath}
\usepackage{latexsym}
\usepackage{longtable}
\begin{document}
\title{Non-Gaussian resistance noise in the ferromagnetic insulating state of a hole doped manganite}
\author{Sudeshna Samanta\footnote[1]{email: sudeshna@bose.res.in} and A. K. Raychaudhuri\footnote[2]{email: arup@bose.res.in,}}
\affiliation{Department of Materials Science, S.N.Bose National Centre for Basic Sciences, Block JD, Sector III, Salt Lake, Kolkata 700 098, West Bengal, India.}
\author{Ya. M. Mukhovskii}
\affiliation{National University of Science and Technology ``MISIS'', Leninsky prosp. 4, Moscow 119049, Russian Federation.}

\date{\today}
\begin{abstract}
\indent
We report the observation of a large $1/f$ noise in the ferromagnetic insulating state (FMI) of a hole doped manganite single crystal of $\mathrm{La_{0.80}Ca_{0.20}MnO_3}$ which manifests hopping conductivity in presence of a Coulomb gap. The temperature dependent noise magnitude shows a deep within the FMI state, there is a sharp freeze out of the noise magnitude with temperature on cooling. As the material enters the FMI state, the noise becomes non-Gaussian as seen through probability density function and second spectra. It is proposed to arise from charge fluctuations in a correlated glassy phase of the polaronic carriers which develop in these systems as reported in recent simulation studies.
\end{abstract}
\maketitle
Electronic transport through localized states in disordered and correlated electronic systems has been a topic of considerable interest\cite{Shlovski,Imry}. In such systems, long range Coulomb interaction can lead to opening-up of a soft gap in the density of states (referred to as Coulomb gap, $\Delta_{CG}$) and hopping conduction in presence of such a gap\cite{Efros1}. Another consequence is emergence of ``glassy'' slow relaxations of charge carriers arising from a large number of low-lying states separated by barriers\cite{Davies}. Such a glassy state can lead to enhanced low frequency ($f$) non-Gaussian resistance noise (typically with power spectrum varying as $1/f$) arising from charge fluctuations\cite{Yu,Burin}. These issues in a Coulomb glass have been reviewed recently\cite{Imry}. The experimental investigations on these questions are few and were carried out only in the doped semiconductors with electron density close to the critical concentration ($n_C$) for metal-insulator (MI) transition\cite{Messy,kar} or in 2DEG in MOSFET's\cite{MOS}. Here, we focus on the issue of non-Gaussian low frequency noise in the Coulomb glass state of a very different material, namely the low hole-doped rare-earth manganites which can have a ferromagnetic insulating (FMI) state below a certain temperature ($T$). In these systems, the Coulomb glass phase occurs for the localized polaronic carriers (in sharp contrast to doped semiconductors) which arise due to strong electron-phonon coupling arising from Jahn-Teller distortion around the Mn$^{3+}$ ions.
\indent

The FMI state of hole doped manganites ($\mathrm{La_{1-x}Ca_{x}MnO_3}$) arises when hole concentration $\mathrm{x}$ in $\mathrm{LaMnO_3}$ exceeds the critical concentration for ferromagnetism $\mathrm{x\approx 0.15}$ yet it is smaller than the concentration needed for the formation of the metallic ground state $\mathrm{x\geq x_{C}=0.22}$. Existence of a Coulomb glass behavior in the FMI state has been inferred from certain transport experiments\cite{jain1,jain2} and also predicted theoretically recently\cite{shenoyPRL}.  The FMI state in manganite is unique\cite{TVRreveiw}. We observe a large low frequency resistance fluctuations (noise) with non-trivial $T$ dependence in the FMI state and the noise becomes non-Gaussian at temperatures below the onset of the insulating state. At low $T$ well below the transition to the insulating state the noise shows a sharp fall on cooling. We propose below that the large noise as well as its $T$ dependence arise from charge fluctuations due to special nature of carriers in it.

\indent
We have done investigation of the low frequency resistance noise ($f<$10Hz) in single crystal of the manganite, $\mathrm{La_{0.80}Ca_{0.20}MnO_3}$ (LCMO20), grown by floating-zone technique\cite{Flzone}. 
For comparison and as a reference system, we have also measured the resistance noise in high quality single crystal of $\mathrm{La_{0.67}Ca_{0.33}MnO_3}$ (LCMO33) which shows a low resistance ferromagnetic metallic (FMM) state. The resistance noise in such a metallic system, as we will show below, is very low and this establishes the noise floor.
(Note: The noise experiments on single crystals allow us to avoid extraneous influences coming from structural defects\cite{ss} which may mask the noise from some of the intrinsic effects like the one being investigated here.)
The noise measurements were carried out on low current biased samples (with evaporated contact pads) using a 5-probe technique\cite{schofield} followed by a series of digital signal processing techniques\cite{akr}.
A complete set of time series of voltage jumps ($|\Delta \rm{V}|$) at each $T$ (stabilized to within $\pm$1mK) consists of 5$\times10^6$ data points or more have been used to obtain the power spectral density $S_{V}(f)$, the second spectrum (defined later on) as well as the probability density function (PDF) (probability $P(|\Delta \rm{V}|)$ for occurrence of the voltage jump $|\Delta \rm{V}|$).
The last two quantities give measure of the non-Gaussian component (NGC) of the fluctuations. The data are taken for 40K$\le T \le$ 300K. Below 40K the resistivity of the material becomes very large for a reliable noise measurement and no heating effect are seen.
\begin{figure}[t]
\begin{center}
\includegraphics[width=8.5cm,height=5.5cm]{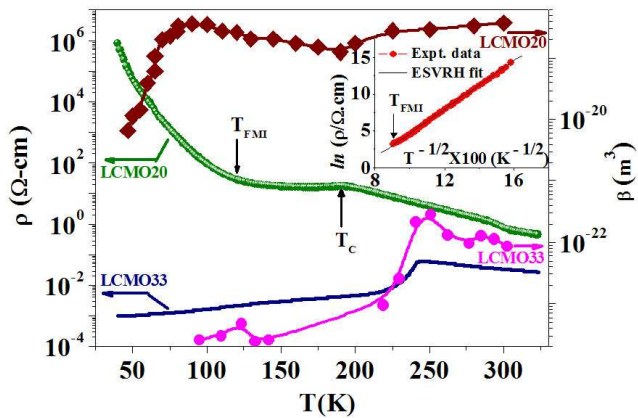}
\end{center}
\caption{(Color online) Variation of $\rho$ (left ordinate) and scaled noise amplitude $\beta$ (right ordinate) as a function of $T$ for LCMO20 and LCMO33. The $T_C$ for both samples and $T_{FMI}$ for LCMO20 are marked. Inset shows $ln\rho$ vs $T^{-1/2}$ plot for LCMO20 for $T\le T_{FMI}$}.
\label{Fig1}
\end{figure}

\indent
In Fig.~\ref{Fig1} we plot the $T$ dependence of the resistivity $\rho(T)$ for LCMO20 along with LCMO33 for comparison. For LCMO20, a paramagnetic-ferromagnetic transition occurs at $T$ = $T_{C}\simeq$ 185K with a shallow peak in $\rho(T)$. (The onset of ferromagnetism has been also established independently through separate magnetic measurements.) Both samples show an activated transport above $T_C$. However, LCMO33 has a stable metallic phase for $T<T_C$. For LCMO20, below $T_{C}$, the temperature dependence is shallow over a limited range followed by an upturn at $T = T_{FMI}\simeq$ 120K as the material enters FMI state. There exist a mixed phase region with FMM phase for $T_{FMI}< T <T_{C}$. In the low temperature phase ($T\leq T_{FMI}$), $\rho(T)$ is an insulator that follows Efros-Shklovskii variable range hopping (ESVRH) relation $\rho = \rho_{0}{exp(T_{0}/T)^{1/2}}$ (see inset of Fig.~\ref{Fig1}). $T_0$ is related to the localization length $\xi$=$2.8e^{2}/(4\pi \epsilon_{0}\epsilon_{1}k_{B}T_{0})$ where $\epsilon_{1}$ is the dielectric constant. The $T$ dependence of $\rho(T)$ points towards the existence of a soft gap in the density of states characterized by the Coulomb gap $\Delta_{CG}$\cite{Shlovski}. From fitted data we obtain $\rho_0=$7.96$\times$ 10$^{-6}$Ohm-cm.K$^{-1/2}$ and $T_0$ = 2.68$\times 10^4$K. Using $T_{0}$, we obtain a localization length $\xi \approx$ 2$\AA{}$, which is of the order of half the unit cell, indicating strong localization of the polaronic carriers.
\begin{figure}[t]
\begin{center}
\includegraphics[width=6.5cm,height=4.5cm]{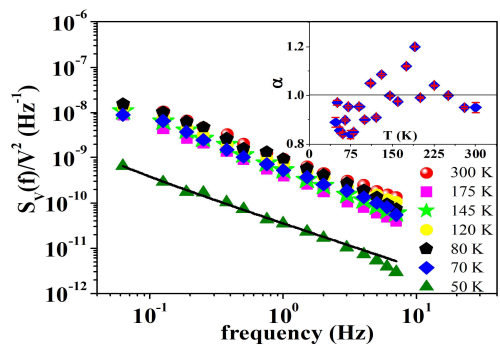}
\end{center}
\caption{(Color online) The normalized spectral power $S_{V}(f)/V^2$ are shown at few representative temperatures as a function of $f$. Inset shows the variation of $\alpha$ with $T$.}
\label{Fig2}
\end{figure}
 
\indent
Specific heat measurements in these materials at low $T$ show a linear term indicating a finite density of states ($N(E_{F})$) at the Fermi-level through localized states\cite{jain1,jain2}. Using $N(E_{F})$ from the experimentally observed linear term, we obtained $\Delta_{CG} \approx$ 150 meV, using the relation $\Delta_{CG}$=$e^{3}N(E_{F})^{1/2}/(4\pi \epsilon_{0}\epsilon_{1})^{3/2}$\cite{Shlovski}. $\Delta_{CG}$ is much larger than in semiconductors which is much smaller than 10meV \cite{Imry}. The large $\Delta_{CG}$ is a direct consequence of large charge carrier density in these materials which is much larger than the critical concentration for MI transition in doped semiconductors.

\indent
In Fig.~\ref{Fig1} we also plot the relative variance of the resistance noise, $\beta=\Omega \frac{<(\Delta R)^2>}{R^2}$, as a function of $T$. The normalized variance $<(\Delta R)^2>/R^2$ $\equiv (1/V^{2})\int^{f_{max}}_{f_{min}}S_{V}(f)df$ ($\Omega$ is the experimental volume of the sample). (A related quantity, Hooge's parameter\cite{Hooge} may be unreliable in this context because an exact knowledge of the charge carrier density in LCMO20 is uncertain.) $\beta$ of LCMO33 and LCMO20 differ quantitatively as well as qualitatively reflecting the difference in the basic electronic nature of the two below $T_C$, one with a FMM ground state and other with a FMI ground state below $T_C$ respectively. The magnitude of $\beta$ for the LCMO33 sets the lower bound of the noise that is expected from structural defects. In the region close to $T_C$ noise has a magnetic origin\cite{ss}. The large noise in LCMO20, as we will discuss below, arises from fluctuations that have non-structural origin. 

\indent
The normalized spectral power $S_{V}(f)/V^2$ are shown at few representative $T$ as a function of a $f$ in Fig.~\ref{Fig2}. The spectral power has a $1/f^{\alpha}$ dependence with $\alpha$ very close to $1$, although a small but interesting variation in the exponent $\alpha$ can be seen close to $T_{FMI}$ as shown in inset of Fig.~\ref{Fig2}. For $T>T_{C}$, $\alpha \geq 1$ and $\alpha \leq 1$ in the FMI state. This is unlike the observation in doped semiconductors close to the MI boundary\cite{kar} where $\alpha \geq 1$.
\begin{figure}[t]
\begin{center}
\includegraphics[width=8.0cm,height=5.0cm]{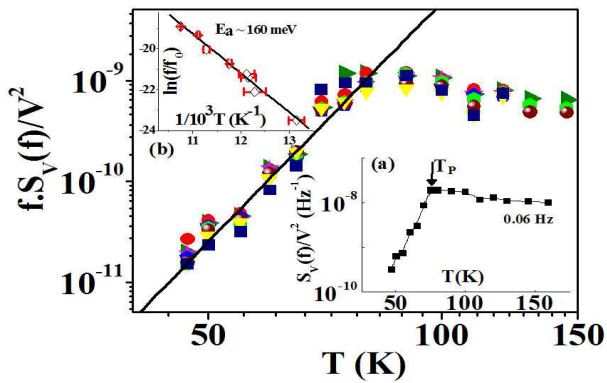}
\end{center}
\caption{(Color online) $f.S_{V}(f)/V^{2}$ as a function of $T$ for few representative $f$ from 0.06 Hz to 10Hz. (a) Variation of $S_{V}(f)/V^{2}$ with $T$ for $f =$ 0.06 Hz showing shallow peak at $T = T_P$ ($< T_{FMI}$) (see text). (b) Arrhenius plot along with the best fit curve of $E_{a}\simeq$160meV.}
\label{Fig3}
\end{figure}

\indent
The $T$ dependence of the spectral power, presented as $f.S_{V}(f)/V^2$, at some representative $f$ from 0.06 Hz to 10Hz are shown in Fig.~\ref{Fig3}. Since $S_{V}(f)\sim 1/f$, $f.S_{V}(f)/V^2$ collapses into almost a single graph for all $f$ over most of the temperature region. Data presented here are limited for $T\le T_{C}$ which is the primary focus of this paper. Below $T_{C}$, $S_{V}(f)$ shows a gradual rise till a shallow peak at $T = T_P$ ($< T_{FMI}$). $T_{P}$ varies approximately from 75K to 90K depending on the measuring $f$. $T_P$ shifts to higher value with higher $f$ following an Arrhenius relation: $f = f_{0} exp(-E_{a} / k_{B}T_{P})$ where $f_{0}\sim 10^{9}$ sec$^{-1}$. Inset in Fig.~\ref{Fig3} shows the Arrhenius plot which gives the best fit value of $E_{a}\simeq$160 meV.

\indent
In Fig.~\ref{Fig3}, below $T_{P}$, the noise power goes down very sharply with $T$ following a power law dependence $f.S_{V}(f)/V^2$ $\propto T^{-\gamma}$, where $\gamma$ is $f$ independent and is $\approx$ 8-9. Fit to the power law is shown as a solid line. The strong $T$ dependence of the spectral power in the regime of activated hopping that occurs below $T_{FMI}$ has not been seen before in manganites. Interestingly, such a sharp drop in $S_{V}(f)$ at low $T$ has been seen in some of the doped semiconductors in hopping regime. In bulk 3D systems, there are only two detailed studies on $T$ dependence of noise spectral power in the hopping regime\cite{Messy,kar}. Very close to the critical concentration of MI transition where the carrier concentration, $n/n_{c}$ $\geq 0.95$, the magnitude of the noise diverges as $T$ is reduced\cite{kar}. However, somewhat away from the critical region, ($n/n_{c}$ $\approx 0.8$), the observed behavior is opposite and the spectral power falls as $T$ decreases. The fall of the spectral power below $T_{FMI}$ thus has similarity with that found in doped semiconductors with carrier concentrations $<0.85$\cite{Messy}. 
\begin{figure}[t]
\begin{center}
\includegraphics[width=6.5cm,height=4.5cm]{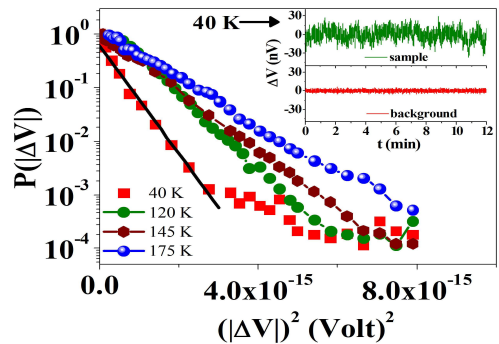}
\end{center}
\caption{(Color online) PDF ($P(|\Delta \rm{V}|)$) as a function of $(|\Delta \rm{V}|)^2$ at few representative $T$. Inset shows time series of the data at 40K for sample as well as for the background for comparison. The background, as can be seen is, has much less power.}
\label{Fig4}
\end{figure}

\indent
In manganites, the electronic states for $T<T_{FMI}$ are expected to be correlated with the development of glassy property according to recent simulation studies\cite{shenoyPRL}. Existence of such correlated state will lead to NGC in the resistance noise in such systems\cite{Yu}. We have investigated the development of such NGC in the fluctuation using two tests, namely, the direct method of the PDF and also the more sensitive second spectrum method. Fig.~\ref{Fig4} shows the PDF, plotted as $ln P(|\Delta \rm{V}|)$ vs ${(|\Delta \rm{V}|)}^2$, for some representative $T$. Example of the observed time series is shown in the inset of Fig.~\ref{Fig4} at 40K. In this plot, a straight line will signify a Gaussian PDF. For $T>T{FMI}$, the PDF has a Gaussian behavior. As $T$ is cooled below $T_{FMI}$, strong deviation from the Gaussian behavior shows up as a long tail for larger values of $|\Delta \rm{V}|$. It is marked for data at $T =$ 40K. This is a direct proof of the appearance of the NGC in the FMI state. 

\begin{figure}[t]
\begin{center}
\includegraphics[width=6.5cm,height=4.5cm]{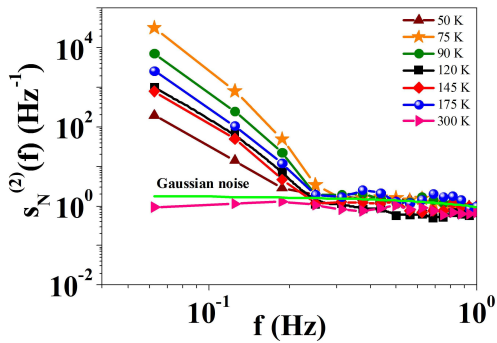}
\end{center}
\caption{(Color online) Normalized second spectra for different $T$. The calculated Gaussian background is plotted as well.}
\label{Fig5}
\end{figure}

\indent
The development of NGC can also be seen from normalized second spectrum estimated from the relation $S^{(2)}_{N}(f) = S^{(2)}(f)/[\int^{f_H}_{f_L}S_{V}(f)/V^{2}]^2$. $S^{(2)}(f) = \int^{\infty}_{0}\langle \Delta v^{2}(t)\Delta v^{2}(t+\tau) \rangle Cos(2\pi f\tau)$ has been calculated within a chosen $f$ band ($f_L$=1Hz, $f_H$=3Hz). For a Gaussian fluctuation $S^{(2)}_{N}(f)$ is unity and its deviation from unity is a measure of non Gaussian fluctuations\cite{SSolin}. In Fig.~\ref{Fig5} we plot $S^{(2)}_{N}(f)$ at few representative $T$. It can be seen that at lower $T$ (and $f<$ 0.2Hz) there is a strong deviation from unity. The observed NGC at low $f$ can be expected if there is a built-up of correlated charge fluctuation with long relaxation times. We take this as a signature of entry into a correlated glassy phase\cite{Yupss}. Our experiment shows that in the FMI state the fluctuation can be large till it freezes at lower $T<<T_{FMI}$ and the fluctuation has a good non-Gaussian contribution, particularly at low $f$. 

\indent
We discuss qualitatively the reason for large noise in the FMI state using a recent model\cite{TVRreveiw} which states that there are two types of carriers in manganites, namely, the localized polaronic carriers (referred as $l$) and band type delocalized carriers (referred as $b$). The simulations based on this model\cite{shenoyPRL} show that the presence of large Coulomb interaction (long range) and disorder (arising from dopant atoms) in doped manganites like LCMO20 can lead to a phase separation in which conducting $l$ carrier majority phase is dispersed with islands of nanoscopic puddles of $b$ phase. The hopping conduction occurs in the majority phase ($l$ phase) that percolates through the sample and develops the classical Coulomb glass regime at low $T$. For FMM phase (LCMO33) the $b$ states are occupied and make the majority phase and contribution of $l$ type carriers to transport is negligible. 
We propose that while the hopping conduction occurs in the $l$ phase, $b$ states will be thermally occupied and will lead to exchange of carriers between the percolating $l$ matrix and $b$ islands leading to large charge fluctuations. If the size of such a polaronic cluster that makes the main backbone of conduction is $N_{l}$ and the variance in fluctuations is $<(\delta N_{l})^{2}>$, then $S_{V}(f)/V^{2} = <(\delta N_{l})^{2}>/N_{l}^2$. Since both the quantities in numerator and denominator have independent $T$ dependences, the resulting $T$ dependence of $S_{V}(f)/V^{2}$ will be decided by the relative strengths. Generally, $N_{l}$ is expected to grow with $T$ with a power law ~\cite{Yupss} and to reach a saturation at higher $T$. $<(\delta N_{l})^{2}>$ is expected to reflect the charge exchange between the $b$ and $l$ regions as well as that between infinite clusters (that is the back bone of conduction) and small clusters within the polaronic glass phase. Below $T_{FMI}$, as the nanoscopic inhomogeneity develops and the $b$ puddles separate out from the $l$ matrix, the $b$-$l$ exchange freezes out. Also the probability of polaronic hopping (that occurs in the Coulomb glass phase of $l$ carriers) becomes exponentially small at lower $T$. These two lead to a rapid drop in $<(\delta N_{l})^{2}>$. In this temperature range though $N_{l}$ will also decrease, but the $<(\delta N_{l})^{2}>$ can have a much stronger dependence leading to sharp drop of the spectral power. The gradual rise of the noise on cooling below $T_C$ reflects the growth of $<(\delta N_{l})^{2}>$ because the $b$ puddles can be occupied and can exchange charges with $l$ phase and also $N_{l}$ will decrease with $T$. The Arrhenius dependence of $f$ and $T_P$ also reflect this activated nature of the fluctuation.

\indent
The observed large NGC can develop due to two reasons. First the coulomb glass phase of polaronic $l$ carriers can lead to non-Gaussian fluctuations\cite{Yupss}. Second, the nanoscopic phase separation can lead to non-Gaussianity due to the mechanism of random distribution of current\cite{seidler}. Both the sources for non-Gaussianity can contribute in tandem due to the nature of the phases present and the transport through them and their relative contributions will also be $T$ dependent.

\indent
In conclusion, the investigation of noise spectroscopy strongly suggests the existence of slow correlated motion of charge carriers in FMI state. The results presented here which have not been observed before add a new insight into the FMI state of manganites. 
The results establish that the FMI state (where the carriers are predominantly polaronic) is a correlated glassy phase, which till now has been found experimentally only in doped semiconductors close to the critical composition of MI transition, MOSFET or in oxides like $\mathrm{In_{x}O_{y}}$\cite{orly}. 
It has been proposed that the temperature dependent relative contributions of the two mechanisms like the nanoscopic phase separation and the glassy response in the polaronic type carriers in LCMO20 can lead to large charge fluctuations leading to large noise with substantial NGC.

\indent
The authors thank the Department of Science and Technology, Govt. of India for financial support.

\end{document}